\documentclass[sigconf,natbib=true]{acmart}
\AtBeginDocument{%
  }

\usepackage{booktabs}   % for \toprule, \midrule, \bottomrule
\usepackage{tabularx}   % for tabularx and automatic column stretching
\usepackage{multirow}
\usepackage{subcaption}
\usepackage{graphicx}
\usepackage{geometry}
\usepackage{microtype}
\usepackage{draftwatermark}

\definecolor{bgtabcolor}{RGB}{247, 210, 173}

\copyrightyear{2026}
\acmYear{2026}
\setcopyright{cc}
\setcctype{by-nc-nd}
\acmConference[RecSys '26]{20th ACM Conference on Recommender Systems}{September 27-October 02, 2026}{Minneapolis, MN, USA}
\acmBooktitle{20th ACM Conference on Recommender Systems (RecSys '26), September 27-October 02, 2026, Minneapolis, MN, USA}
\acmDOI{10.1145/3773078.3831852}
\acmISBN{979-8-4007-2284-4/2026/09}

\newcommand{\commentAP}[1]{\textcolor{black}{}}

\begin{document}
\SetWatermarkText{Preprint}

\title[Binge Watch: Reproducible Multimodal Benchmark Datasets\\for Large-Scale Movie Recommendation on MovieLens-10M and 20M]{\textit{Binge Watch:} Reproducible Multimodal Benchmark Datasets for Large-Scale Movie Recommendation on MovieLens-10M and 20M}
%\title[Binge Watch: Reproducible Multimodal Benchmark Datasets for\\Large-Scale Movie Recommendation on MovieLens-10M and 20M]{\textit{Binge Watch:} Reproducible Multimodal Benchmark Datasets for Large-Scale Movie Recommendation on MovieLens-10M and 20M}

\author{Giuseppe Spillo}
\email{giuseppe.spillo@uniba.it}
\orcid{0000-0001-8345-4232}
\affiliation{%
  \institution{University of Bari Aldo Moro}
  \city{Bari}
  \country{Italy}}

\author{Alessandro Petruzzelli}
\email{alessandro.petruzzelli@uniba.it}
\orcid{0009-0008-2880-6715}
\affiliation{%
  \institution{University of Bari Aldo Moro}
  \city{Bari}
  \country{Italy}}

\author{Cataldo Musto}
\email{cataldo.musto@uniba.it}
\orcid{0000-0001-6089-928X}
\affiliation{%
  \institution{University of Bari Aldo Moro}
  \city{Bari}
  \country{Italy}}

\author{Marco de Gemmis}
\email{marco.degemmis@uniba.it}
\orcid{0000-0002-2007-9559}
\affiliation{%
  \institution{University of Bari Aldo Moro}
  \city{Bari}
  \country{Italy}}

\author{Pasquale Lops}
\email{pasquale.lops@uniba.it}
\orcid{0000-0002-6866-9451}
\affiliation{%
  \institution{University of Bari Aldo Moro}
  \city{Bari}
  \country{Italy}}

\author{Giovanni Semeraro}
\email{giovanni.semeraro@uniba.it}
\orcid{0000-0001-6883-1853}
\affiliation{%
  \institution{University of Bari Aldo Moro}
  \city{Bari}
  \country{Italy}}

\renewcommand{\shortauthors}{Spillo et al.}

\begin{abstract}

As Multimodal Recommender Systems gain interest, high-quality datasets with multimedia side information have become essential. However, much of the current literature reports experiments that rely on small-scale, undocumented, or non-public datasets. In this paper, we introduce \textsc{M$^3$L-10M} and \textsc{M$^3$L-20M}, two large-scale, fully documented and reproducible datasets that enrich MovieLens-10M and MovieLens-20M with multimodal features. Following a documented pipeline, we collect movie plots, posters, and trailers and extract features using state-of-the-art encoders. We publicly release raw data mappings, extracted features, and complete datasets to foster reproducibility and advance the field. Qualitative and quantitative analyses demonstrate the quality of our datasets across multiple perspectives. This work establishes a foundational resource for large-scale, multimodal movie recommendation. Our resource is available at: \url{https://zenodo.org/records/18499145}, with source code at \url{https://github.com/giuspillo/M3L_10M_20M}.

\end{abstract}




\maketitle

\section{Introduction and Motivations}

In recent years, Multimodal Recommender Systems (MRSs)~\cite{liu2024multimodal} have emerged as a prominent research direction, exploiting multiple modalities such as images, text, audio, and video to capture richer semantic representations and improve performance in sparse settings~\cite{liu2024multimodalpretraining}. However, empirical evaluation is constrained by the limited availability of large-scale and reproducible datasets. While MovieLens variants~\cite{harper2015movielens} are among the most widely used benchmarks in RS research~\cite{spillo2025see, zhou2023comprehensive, anelli2018knowledge}, their adoption in MRSs has been limited and inconsistent. 
Existing studies often rely on non-publicly available versions~\cite{wei2019mmgcn,lei2023learning}, heavily post-processed variants with undocumented preprocessing~\cite{yi2022multi}, or small-scale versions (ML-100K, -1M) that prevent a full assessment of multimodal features in large-scale scenarios~\cite{attimonelli2025large, spillo2026mmgcf, spillo2026gotta}. Specifically, ML-10M is often significantly altered via aggressive filtering~\cite{tao2020mgat, wei2020graph, liu2022elimrec, du2022invariant, sun2020multi, chen2022breaking,wang2021dualgnn,tao2022self}.
ML-20M remains largely overlooked, with the exception of CineFinder~\cite{saricciccekcinefinder}, though it ignores movie trailers and does not release either the extracted features or the extraction scripts.

These issues highlight a gap between the increasing complexity of models and the need for realistic, large-scale benchmarking. The lack of standardized datasets and transparent protocols prevents full reproducibility, thus limiting research advancement. In this paper, we address these gaps by enriching MovieLens-10M and MovieLens-20M with multimodal features extracted from movie plots (textual), movie posters (images), and movie trailers (acoustic and video). We call these datasets \textbf{\underline{M}}ulti\textbf{\underline{M}}odal \textbf{\underline{M}}ovie\textbf{\underline{L}}ens (M$^3$L). Our datasets preserve the original user–item interactions, avoid heavy core filtering, and provide the highest possible item coverage. 

To sum up, we provide the following contributions: \textit{(i)} we release M$^3$L-10M and M$^3$L-20M datasets; \textit{(ii)} we publicly release the mappings with URLs to access raw data files to foster reproducibility; \textit{(iii)} we encode data with SOTA encoders and release the ready-to-use extracted features, thus reducing the cost of feature extraction; and \textit{(iv)} we perform qualitative and quantitative analysis on the multimodal features. The remainder of this paper discusses our pipeline to enrich ML-10M and ML-20M (Section \ref{sec:metho}), followed by our qualitative and quantitative analyses (Sections \ref{sec:qual_exp} and \ref{sec:qunt_exp}).

\section{M$^3$L: \underline{M}ulti\underline{M}odal \underline{M}ovie\underline{L}ens Datasets}\label{sec:metho}

In this Section, we describe our resource. First, we introduce the pipeline to enrich MovieLens datasets with multimodal information. Then, we focus on the encoders used to extract multimodal features.

\subsection{Multimodal Raw Data Collection}

The process discussed below is specifically designed for the ML-20M dataset. However, ML-10M is a subset of ML-20M, so the pipeline can be applied to the smaller dataset, or, equivalently, ML-10M can even be directly extracted from the larger one.

\noindent{\textbf{(1) Accessing MovieLens Metadata.}} 
Our data collection pipeline starts from the original MovieLens dataset, which can be accessed on the GroupLens website\footnote{\url{https://grouplens.org/datasets/movielens/}}. 
MovieLens data includes user ratings and movie metadata (movie titles, genres, and textual user tags). 
Each movie also includes a YouTube link. However, many of these links are now inaccessible due to the removal of movie trailers.
In addition, each movie is provided with its Internet Movie Database\footnote {\url{https://www.imdb.com/it/}} (IMDb) and The Movie Database\footnote {\url{https://www.themoviedb.org/}} (TMDB) identifier number. 

\noindent{\textbf{(2) Querying TMDB.}}
In our pipeline, we focus on TMDB identifiers, since the platform provides APIs that allow to collect raw multimodal side information, such as text, images, and video, describing the items. To do so, we first obtain a TMDB API key\footnote{The API key is free for research purposes.}.
Then, we set up a script to query TMDB using movie IDs and to obtain, for each movie, its title, plot, and links to the movie poster and trailer.  However, it is worth noting that some pieces of information are not available for all movies. Some statistics about the coverage of the data extraction process are discussed next.  %Unfortunately, not all movies include all three pieces of information. However, we continue to query and collect available information. Also, for this step, we release the script for querying TMDB from our repository.

\noindent{\textbf{(3) Downloading Raw Data.}}
Next, based on the URLs obtained in the previous step by querying TMDB, we download the posters and trailers for each movie using the Python libraries \texttt{requests} and \texttt{yt\_dlp}, respectively.
If the trailer is unavailable due to regional licensing constraints or because the associated YouTube videos are set to private, we use the YouTube link in the original GroupLens and download the trailer. If this trailer is not available as well, we remove the movie from the dataset. 
It is important to note that once raw data is downloaded, we encode it using multimodal encoders and then delete the raw files. Moreover, we do not release the raw files to avoid copyright infringement\footnote{To ensure copyright compliance, our pipeline uses a non-consumptive framework. We avoid redistributing raw streams and provide only YouTube IDs and transformative embeddings. As mathematical abstractions, these latent representations cannot reconstruct original media, ensuring they don't substitute for the creative work. This aligns with 'Fair Use' (17 U.S.C. § 107) for non-commercial research—preserving creator rights while enabling scientific reproducibility.}. \commentAP{We instead provide pre-computed embeddings, which aligns with standard practice since most multimodal recommendation architectures use offline-extracted features as input~\cite{yi2022multi, spillo2025see, yi2025enhancing, attimonelli2025large, zhou2025doesmultimodalityimproverecommender, wei2019mmgcn}.}

We temporarily cache the raw data from posters and videos before extracting the image, audio, and video features.
As shown in Table \ref{tab:stats_raw_data}, after the download step, we observe a high coverage of movie plots and movie posters for both ML-10M and ML-20M, and a very good coverage for movie trailers for ML-20M. 
The lower coverage of movie trailers is mainly due to geographical restrictions. Moreover, other videos are associated with deleted accounts or have been removed from the platform due to copyright issues.
However, although some items are missing, the number of items for which \textit{all} multimodal data is available is significantly higher than in previous works (see Table \ref{tab:dataset_comparison_enriched}). 
In our repository, we release all the scripts for downloading raw data from the original MovieLens metadata. Our resource is also fully accessible on Zenodo\footnote{\url{https://zenodo.org/records/18499145}}.

\begin{table}[h]
\centering
\caption{Multimodal Item Raw Data Coverage per dataset.}
\label{tab:stats_raw_data}
\resizebox{0.35\textwidth}{!}{
\begin{tabular}{l cc cc}
\toprule
& \multicolumn{2}{c}{\textbf{ML-10M}} & \multicolumn{2}{c}{\textbf{ML-20M}} \\
& \multicolumn{2}{c}{($N=10,677$)} & \multicolumn{2}{c}{($N=26,744$)} \\
\cmidrule(lr){2-3} \cmidrule(lr){4-5}
\textbf{Modality} & \textbf{Coverage} & \textbf{\%} & \textbf{Coverage} & \textbf{\%} \\
\midrule
Plots    & 10,328 & 96.73 & 24,713 & 92.41 \\
Posters  & 10,304 & 96.51 & 24,580 & 91.91 \\
Trailers & 9,072  & 84.97 & 19,142 & 71.57 \\ 
\midrule
\textbf{Overlap}  & 9,031  & 84.58 & 19,009 & 71.08 \\
\bottomrule
\end{tabular}
}
\end{table}

\begin{table}[h]
\centering
\caption{Comparison of Dataset Statistics: ML vs. our M$^3$L}
\label{tab:stats_datasets}
\resizebox{0.48\textwidth}{!}{
\begin{tabular}{l ccc c ccc}
\toprule
 & \multicolumn{3}{c}{\textbf{ML-10M}} & & \multicolumn{3}{c}{\textbf{ML-20M}} \\
\cmidrule{2-4} \cmidrule{6-8}
\textbf{Metric} & \textbf{Original} & \textbf{M$^3$L} & \textbf{\%} & & \textbf{Original} & \textbf{M$^3$L} & \textbf{\%} \\
\midrule
\# Users   & 69,878     & 69,878     & 100.0\%  & & 138,493    & 138,493    & 100.0\% \\
\# Items   & 10,677     & 9,031      & 84.58\%  & & 26,744     & 19,009     & 71.08\% \\
\# Ratings & 10,000,054 & 9,409,884  & 94.10\%  & & 20,000,263 & 18,777,965 & 93.89\% \\
Sparsity   & 98.66\%    & 98.51\%    & --      & & 99.46\%    & 99.29\%    & --     \\
\bottomrule
\end{tabular}
}
\end{table}

\subsection{Feature Extraction}
After acquiring multimodal raw data, we extract fixed-size embeddings to provide a unified, reproducible representation of items across modalities.
To this end, we encode \textit{(i)} plot summaries (text), \textit{(ii)} posters (images), and \textit{(iii)} trailers (video and audio) by exploiting pretrained state-of-the-art encoders widely adopted in previous works in the multimodal RS field~\cite{yi2022multi, spillo2025see, yi2025enhancing, attimonelli2025large, zhou2025doesmultimodalityimproverecommender, wei2019mmgcn}. The list of encoders is presented below.

\noindent\textbf{(1) Textual modality.} Plots are encoded using sentence-level transformers: {MiniLM}~\cite{MINILM}, a compact distilled model (\texttt{all-\\MiniLM-L6-v2}); {MPNet}~\cite{song2020mpnet}, which captures bidirectional context (\texttt{all-mpnet-base-v2}); and {CLIP-Text}~\cite{radford2021clip}, aligning text and images in a shared embedding space (\texttt{clip-ViT-B-32}).

\noindent\textbf{(2) Visual modality.} Posters are encoded with {VGG16}~\cite{simonyan2014verydeep}, a 16-layer CNN backbone; {Vision Transformer (ViT)}~\cite{dosovitskiy2021image}, modeling global context via image patches (\texttt{vit-base-patch16-224-in21k}); and {CLIP-Image}~\cite{radford2021clip}, a multimodal model trained with contrastive learning (\texttt{clip-ViT-B-32}).

\noindent\textbf{(3) Video modality.} Spatio-temporal trailer features are extracted via {SlowFast (R50)}~\cite{feichtenhofer2019slowfast} (dual-pathway dynamics), {R(2+1)D}~\cite{tran2018closer} (factorized 3D CNN), and {MViT}~\cite{fan2021mvit} (multiscale vision transformer).

\noindent\textbf{(4) Audio modality.} Trailers are processed (16 kHz mono) and encoded using {VGGish}~\cite{hershey2017cnn} (YouTube-trained CNN), {Whisper}~\cite{radford2022whisper} (using \texttt{whisper-base} encoder for the first 30 seconds), and {AST}~\cite{gong2021ast} (Audio Spectrogram Transformer operating on log-mel patches).

In our repository, we publicly release the embeddings, enabling fair comparisons across models and minimizing implementation variance. Once we obtain all the embeddings, we then compute the number of items for which all modalities are available.
This value, reported in Table \ref{tab:stats_datasets}, serves as the starting point to filter out the original ML-10M and ML-20M to obtain our M$^3$L-10M and M$^3$L-10M datasets. As we can observe, we retained the majority of the interaction data from the original datasets (\textit{i.e.}, more than 90\%). 

In addition, in Table \ref{tab:dataset_comparison_enriched}, we compare our M$^3$L-10M with the versions of ML-10M previously adopted in the literature on MRSs.
In particular, we report the resulting dataset sizes (\#users, \#items, \#ratings, \#ratings/user) after preprocessing, along with the set of available modalities (T=text, I=image, A=audio, V=video). \textbf{To conclude, our datasets offer greater coverage and richer side information than prior ML-10M variants~\cite{wei2019mmgcn, yi2022multi}, combining larger user–item interactions with a wider spectrum of modalities (text, image, audio, video). Moreover, our pipeline is fully reproducible.}

\begin{table}[h]
\centering
\caption{Comparison of Our ML-10M against other ML-10M variants used in MMGCN's~\cite{wei2019mmgcn} and MMGCL's~\cite{yi2022multi}.}
\label{tab:dataset_comparison_enriched}
\resizebox{0.45\textwidth}{!}{% Resize to fit width if needed
\begin{tabular}{l c c c c c}
\toprule
 & \textbf{Our M$^3$L-10M} & \textbf{Used in~\cite{wei2019mmgcn}} & \textbf{\%} & \textbf{Used in~\cite{yi2022multi}} & \textbf{\%} \\
\midrule
\# users & 69,878 & 55,485 & 79.4\% & 12,674 & 18.1\% \\
\# items & 9,031 & 5,986 & 66.3\% & 4,214 & 46.7\% \\
\# ratings & 9,409,884 & 1,239,508 & 13.2\% & 1,013,573 & 10.8\% \\ \midrule
\# rat./user & $\sim$135 & $\sim$22 & - & $\sim$80 & - \\ 
\# rat./item & $\sim$1,042 & $\sim$207 & - & $\sim$241 & - \\  \midrule
Modalities & T, I, A, V &  T, I, A & - & T, I, A & -  \\
\bottomrule
\end{tabular}
}
\end{table}

% \vspace{-10px}
\section{Qualitative Analysis}\label{sec:qual_exp}

\begin{figure*}[htbp]
    \centering
    % --- Subfigure 1: Text ---
    \begin{subfigure}[b]{0.24\textwidth}
        \centering
        \includegraphics[width=\textwidth]{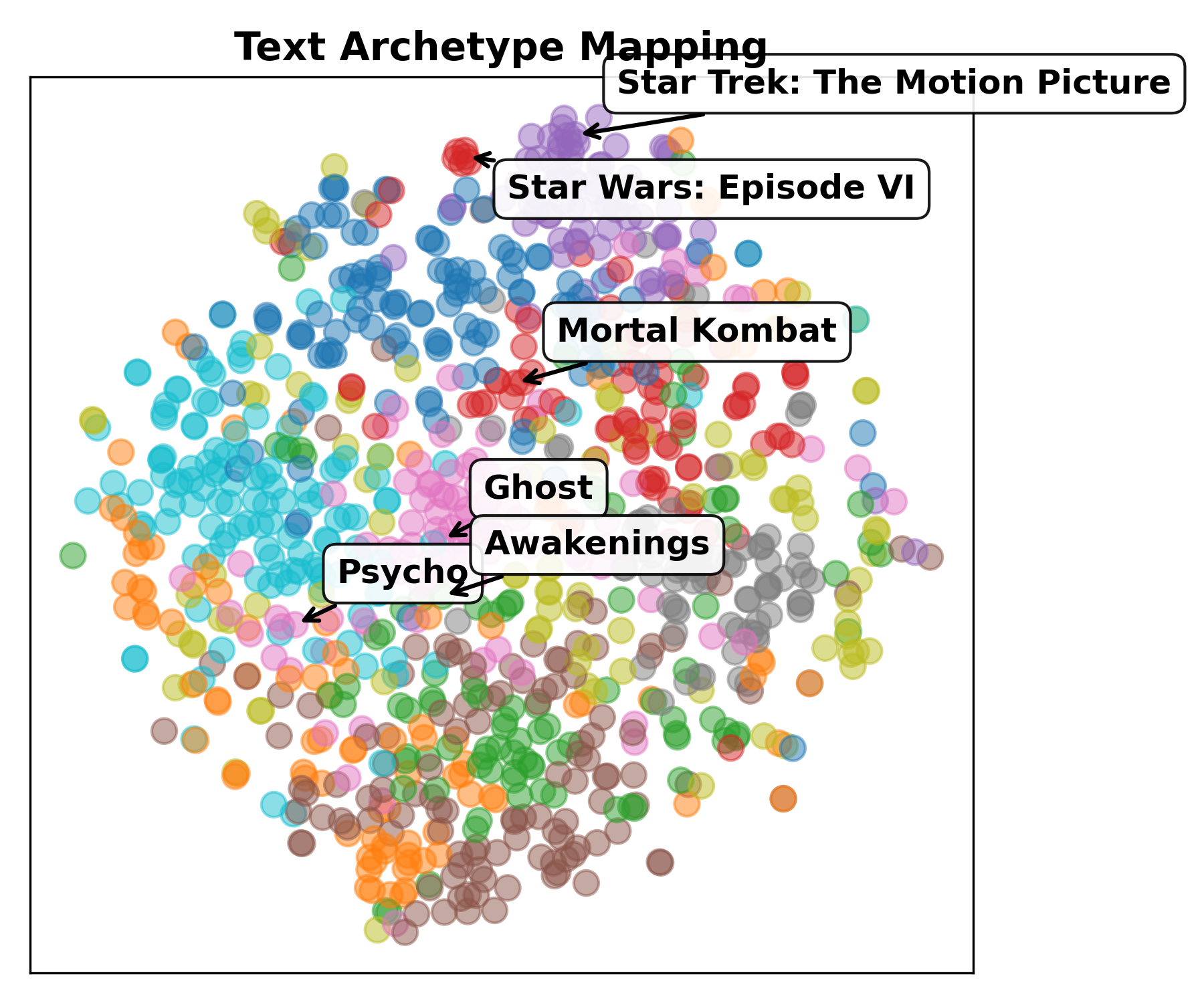}
        \caption{Text (MPNet)}
        \label{fig:tsne_text}
    \end{subfigure}
    \hfill
    % --- Subfigure     2: Image ---
    \begin{subfigure}[b]{0.24\textwidth}
        \centering
        \includegraphics[width=\textwidth]{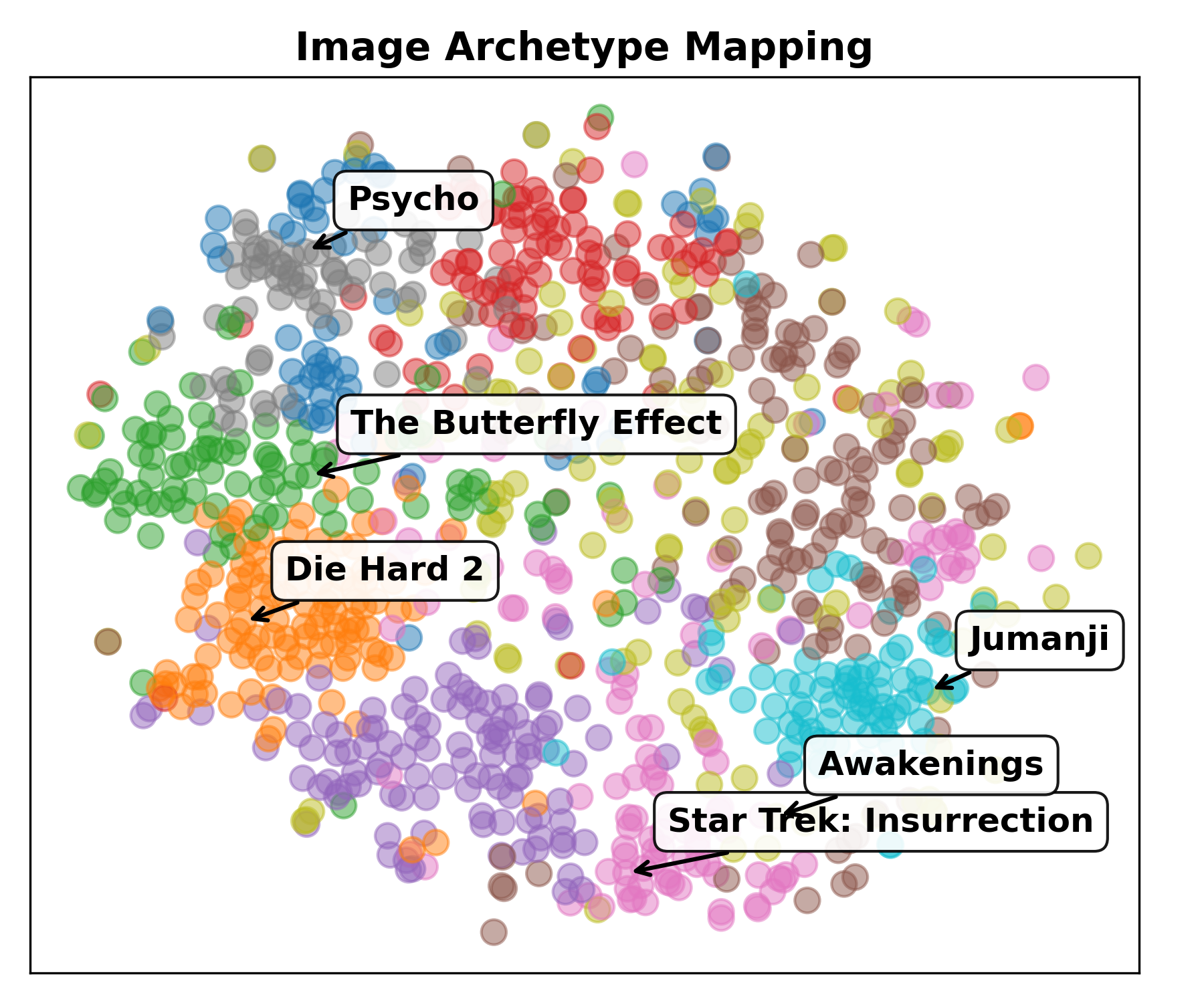}
        \caption{Image (ViT)}
        \label{fig:tsne_image}
    \end{subfigure}
    \hfill
    % --- Subfigure 3: Video ---
    \begin{subfigure}[b]{0.24\textwidth}
        \centering
        \includegraphics[width=\textwidth]{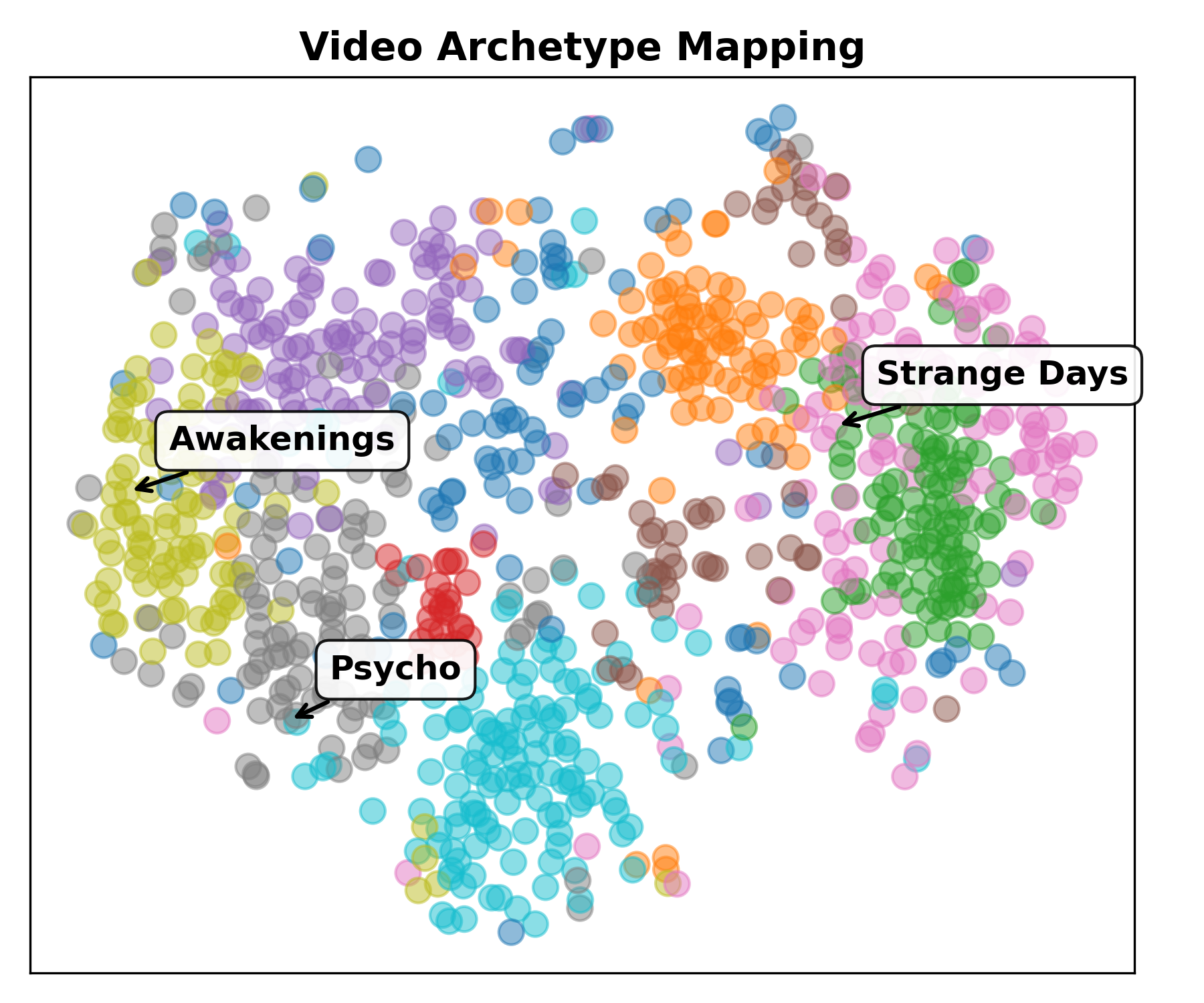}
        \caption{Video (SlowFast)}
        \label{fig:tsne_video}
    \end{subfigure}
    \hfill
    % --- Subfigure 4: Audio ---
    \begin{subfigure}[b]{0.24\textwidth}
        \centering
        \includegraphics[width=\textwidth]{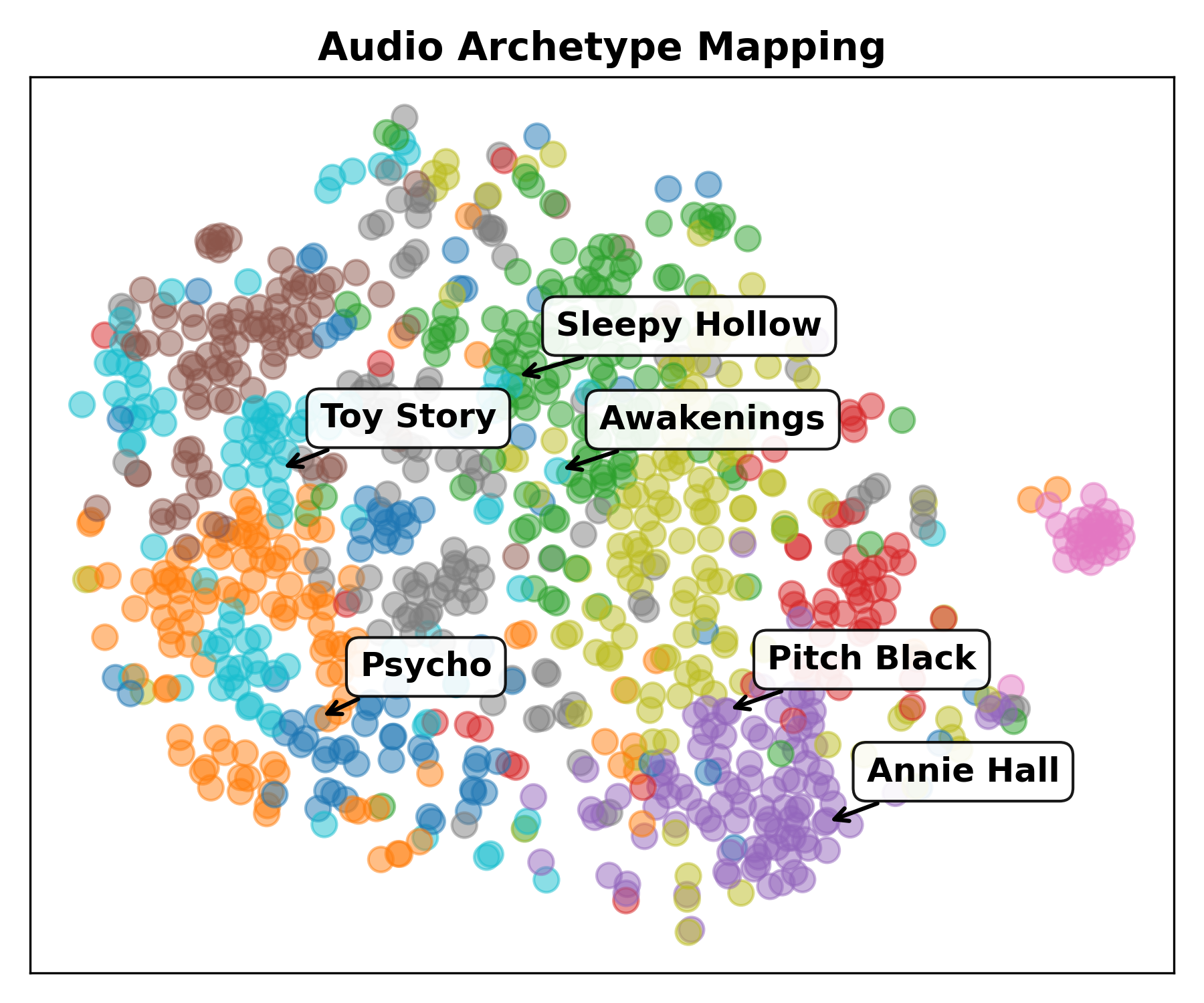}
        \caption{Audio (AST)}
        \label{fig:tsne_audio}
    \end{subfigure}
    
    \caption{t-SNE projections revealing structural differences across modalities. Text forms dense narrative clusters; \textit{Psycho} and \textit{Awakenings} overlap at the center due to shared psychological themes. Image exhibits spatial separation through visual archetypes, placing both films in distinct regions that reflect different color palettes and cinematography. Video and Audio display greater dispersal of kinetic and atmospheric properties; here, \textit{Psycho} and \textit{Awakenings} co-locate in Video space via shared pacing, demonstrating how stylistic modalities encode non-narrative cinematic properties that transcend genre and era.}
    \label{fig:tsne_clusters}
\end{figure*}

To characterize the semantic properties of the different modalities, we carry out two qualitative analyses. First, we use t-SNE to visualize the latent spaces (Figure~\ref{fig:tsne_clusters}), then we exploit a radar chart to \textit{fingerprint} the characteristics of each modality (Figure~\ref{fig:radar_chart}).

\subsection{Multimodal Latent Space Distributions}

To visually inspect the semantic properties of the extracted features, we use t-SNE projections of the high-dimensional embeddings across all modalities, as shown in Figure~\ref{fig:tsne_clusters}.

The \textbf{Text} modality (Fig.\ref{fig:tsne_text}) shows dense structural concentration. Proximity between \textit{Star Wars: Episode VI} and \textit{Star Trek: The Motion Picture} validates the captured plot similarities; these, along with the separation that emerged between action movies like \textit{Mortal Kombat} and psychological movies such as \textit{Ghost}, confirm that the use of text forms regions that align with traditional genre classifications.

\textbf{Image} (Fig.~\ref{fig:tsne_image}) acts as a stylistic middle ground; archetypes like \textit{Die Hard 2} and \textit{Star Trek: Insurrection} occupy lobes defined by color and lighting. This shows that while Text provides thematic anchors, stylistic modalities offer essential, non-redundant signals that offer a different perspective on the item space.

\textbf{Video} and \textbf{Audio} (Figs.~\ref{fig:tsne_video}, \ref{fig:tsne_audio}) capturing kinetic/atmospheric properties independent of narrative. This leads to greater spatial dispersion among clusters. For instance, the video space co-locates \textit{Psycho} and \textit{Awakenings} due to shared visual stasis, while the high-motion \textit{Strange Days} is projected at the opposite extreme.

\subsection{Modality Fingerprinting}
\begin{figure}[htbp]
      \centering
      \includegraphics[width=0.3\textwidth]{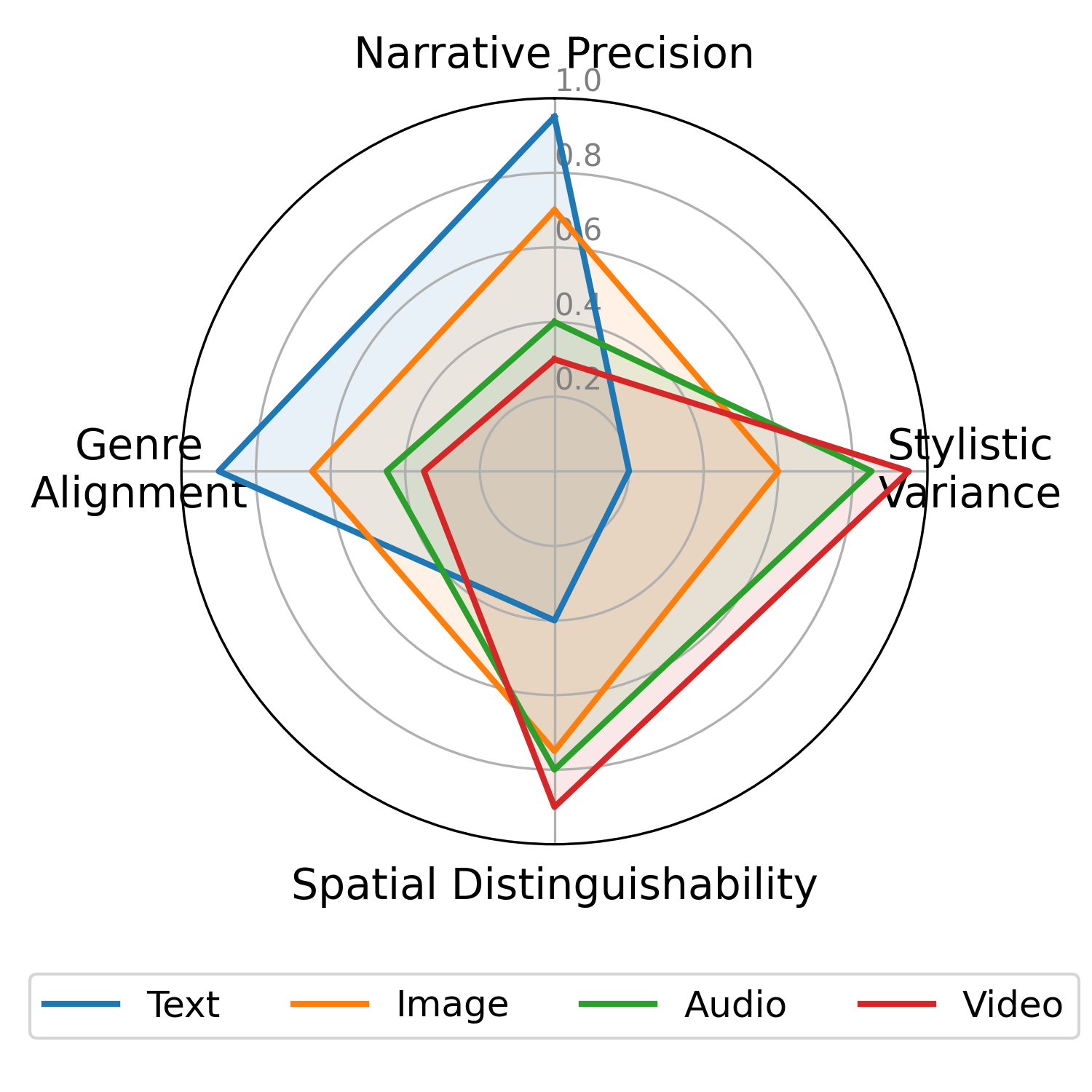}
          % \vspace{-10pt}
\caption{Modality Fingerprinting: Comparison of thematic versus stylistic profiles across normalized dimensions.}
    \label{fig:radar_chart}
    % \vspace{-5pt}
\end{figure}

To characterize the contribution of each modality, we use a radar chart to compare the extracted features across four core dimensions: \textit{Narrative Precision}, \textit{Stylistic Variance}, \textit{Spatial Distinguishability}, and \textit{Genre Alignment} (Figure \ref{fig:radar_chart}). These dimensions are derived by mapping ground-truth metadata available in MovieLens to latent spaces. As metadata, we used the MovieLens \textit{genome scores}, a dataset that provides the association, expressed as relevance scores, between each movie and a set of descriptive tags.

To define the dimensions, we classify the available genome tags into two categories: \textbf{narrative tags}, which capture plot-driven and thematic elements such as "story," "dialogue," "plot complexity," and "character development," and \textbf{stylistic tags}, which represent cinematic execution elements such as "cinematography," "atmosphere," "pacing," "visual effects," and "soundscape." This manual classification enables us to assess whether each modality's latent space aligns more strongly with narrative content or stylistic execution. The scripts for this analysis are available in our repository.
Next, we compute the four fingerprinting dimensions as follows:
\begin{itemize}
    % \item \noindent\textbf{Narrative Precision:} This dimension quantifies a modality's ability to distinguish plot-driven themes by measuring the inter-cluster variance of narrative tag relevance scores.  Higher variance indicates that the modality creates clusters with distinct narrative profiles. The \textbf{Text} modality demonstrates superior precision ($0.95$), whereas \textbf{Video} scores significantly lower ($0.30$) due to its focus on kinetic patterns rather than explicit storytelling.

    \item \noindent\textbf{Narrative Precision:} Measures plot-driven distinction via inter-cluster variance of \textit{narrative tags}; higher values denote distinct narrative profiles. \textbf{Text} shows high variance ($0.95$), while \textbf{Video} scores lower ($0.30$) by prioritizing visual patterns over explicit storytelling expressed through text.
    
    % \item \noindent\textbf{Stylistic Variance:} This metric captures the diversity of cinematic execution by measuring the inter-cluster variance of stylistic tag relevance scores. A high variance indicates that the modality successfully differentiates between items with distinct stylistic signatures (e.g., fast-paced action vs. atmospheric slow cinema). \textbf{Video} ($0.95$) and \textbf{Audio} ($0.85$) dominate this dimension, identifying stylistic elements that vary substantially even among items within the same genre.

    \item \noindent\textbf{Stylistic Variance:} Measures cinematic diversity via inter-cluster variance of \textit{stylistic} tags; high values differentiate distinct signatures (e.g., fast action vs. slow cinema). \textbf{Video} ($0.95$) and \textbf{Audio} ($0.85$) lead this dimension, identifying stylistic elements that vary even within the same genre.
    
    % \item \noindent\textbf{Spatial Distinguishability:} This refers to the structural separation of data points within the high-dimensional latent space, quantified using the \textit{Silhouette Coefficient}, which measures how well-separated the clusters are (values range from 0 to 1). While Text forms a semantically precise but densely concentrated central mass, \textbf{Video} ($0.90$) and \textbf{Audio} ($0.80$) generate well-separated clusters for items representing stylistic extremes.

    \item \noindent\textbf{Spatial Distinguishability:} Quantifies latent space structural separation via the \textit{Silhouette Coefficient} (0--1). While \textbf{Text} forms a dense central mass, \textbf{Video} ($0.90$) and \textbf{Audio} ($0.80$) create well-separated clusters representing stylistic extremes.
    
    % \item \noindent\textbf{Genre Alignment:} This indicates the degree of overlap with traditional genre taxonomies provided by MovieLens (e.g., "Action," "Horror," "Comedy"), calculated using Adjusted Mutual Information between the latent feature clusters and the original genre labels. Textual features exhibit the highest alignment ($0.90$), confirming they follow traditional genre boundaries, whereas Video pacing patterns ($0.35$) often bridge multiple genres.

    \item \noindent\textbf{Genre Alignment:} Measures overlap with MovieLens genre taxonomies using Adjusted Mutual Information between latent clusters and labels. Textual features show strong alignment ($0.90$) with traditional genre boundaries, whereas Video pacing ($0.35$) often bridges multiple genres.
\end{itemize}

This fingerprinting analysis reveals that while Text provides the thematic \textit{"what"}, the stylistic modalities (Audio and Video) provide the cinematic \textit{"how"}. The inclusion of these stylistic signals introduces new information, such as pacing and atmosphere, which are essential for a holistic understanding of the item space and for maximizing recommendation performance.

% \vspace{-5pt}

\section{Quantitative Analysis}\label{sec:qunt_exp}

To assess the quality and utility of the collected features, we conduct a preliminary quantitative analysis on M$^3$L-10M. Our experiments serve as a foundational validation of the datasets' relevance rather than an exhaustive benchmark. Using the MMRec framework~\cite{zhou2023mmrec}, we compare three representative multimodal models (VBPR~\cite{he2016vbpr}, LATTICE~\cite{zhang2021mining}, FREEDOM~\cite{zhou2023tale}) and a standard BPR~\cite{rendle2012bpr} baseline. 

Our protocol follows a two-stage hierarchical approach, as in~\cite{spillo2025see}: we first evaluate individual modalities to identify the best-performing encoders, and then perform a comparative analysis across all possible modality pairs. Due to space reasons, we only report the results obtained with multiple modalities.
The dataset is split 8:1:1, and models are trained for at most $1000$ epochs with early stopping on Recall@20. While we evaluate performance using Precision, Recall, NDCG, and MAP at $k \in \{5, 10, 20, 50\}$, we report only Recall@10 and NDCG@10 due to space constraints. Full results and hyperparameter details are available in our repository.

\begin{table}[h]
\centering
\caption{Quantitative Analysis Results}
\label{tab:quantitative_exp}
\resizebox{0.45\textwidth}{!}{% Resize table to fit page width if necessary
\begin{tabular}{c c c c c}
\toprule
\textbf{Model} & \textbf{Modalities} & \textbf{Encoders} & \textbf{Recall@10} & \textbf{NDCG@10} \\
\midrule
BPR & / & / & 0.1911 & 0.2142 \\
\midrule \midrule
\multirow{6}{*}{VBPR} & Text + Image & MPNet + CLIP & 0.1897 & 0.2124 \\
& Text + Audio & MPNet + Whisper & \textbf{\colorbox{bgtabcolor}{0.1938}} & \textbf{\colorbox{bgtabcolor}{0.2165}} \\
& Text + Video & MPNet + MViT & 0.1919 & 0.2139 \\
& Image + Audio & CLIP + Whisper & 0.1893 & 0.211 \\
& Image + Video & CLIP + MViT & 0.1862 & 0.2084 \\
& Audio + Video & Whisper + MViT & 0.1906 & 0.2124 \\
\midrule
\multirow{6}{*}{LATTICE} & Text + Image & CLIP + VGG & 0.1746 & 0.1919 \\
& Text + Audio & CLIP + AST & 0.1723 & 0.1842 \\
& Text + Video & CLIP + MViT & 0.1741 & 0.1886 \\
& Image + Audio & VGG + AST & \textbf{0.1764} & \textbf{0.1977} \\
& Image + Video & VGG + MViT & 0.1763 & 0.1952 \\
& Audio + Video & AST + MViT & 0.1744 & 0.1911 \\
\midrule
\multirow{6}{*}{FREEDOM} & Text + Image & MPNet + VGG & 0.185 & 0.2126 \\
& Text + Audio & MPNet + VGGish & 0.1866 & 0.2138 \\
& Text + Video & MPNet + MViT & 0.1832 & 0.21 \\
& Image + Audio & VGG + VGGish & \textbf{0.1873} & \textbf{0.2146} \\
& Image + Video & VGG + MViT & 0.1848 & 0.2125 \\
& Audio + Video & VGGish + MViT & 0.187 & 0.214 \\
\bottomrule \bottomrule
\end{tabular}
}
\end{table}

{\textbf{Results Discussion.}}
In Table \ref{tab:quantitative_exp}, we analyze the performance of the multimodal RS models in the uni-modal and double-modal settings. Regarding the single-modality experiments, the results highlight the quality of specific feature sets; for instance, the acoustic features (VGGish) enable FREEDOM to achieve its highest single-modality Recall, suggesting significant semantic richness in the acoustic data. The effectiveness of the multimodal features becomes even more evident in the double-modality settings. A consistent pattern emerges across all architectures: the \textit{Audio} modality is a key component of the best-performing configuration for every model. This finding empirically validates the importance of the acoustic features provided in our resource, demonstrating that they are essential for maximizing recommendation performance.
It is worth noting that the best overall result is obtained by VBPR when textual and acoustic features are used together, followed by the configuration that exploits textual and video features. This is a very interesting result, since our qualitative analysis showed that these two modalities encode complementary information (Figure \ref{fig:radar_chart}). This result further suggests the need to consider different modalities that are typically overlooked in the literature (video and audio), which are limited to textual and visual~\cite{zhang2021mining, zhou2023tale, attimonelli2025large, zhou2023mmrec}.
Finally, we observe that BPR achieves good performance, confirming the robustness of the interaction signal in the dataset. While no individual modality is sufficient to outperform the BPR baseline (likely due to low sparsity) the double-modality configuration of VBPR (MPNet + Whisper) successfully overcomes it. 
This confirms that multimodal features provide complementary data to the interaction graph.
% This confirms that the extracted features provide complementary information that effectively augments the interaction graph.

% % \vspace{-5pt}
\section{Conclusions}\label{sec:concl}

In this paper, we presented M$^3$L, a MovieLens-based dataset enriched with multimodal features. Our resources focus on ML-10M and ML-20M, and aim at fostering research in multimodal large-scale movie recommendation. 
We provide a fully documented, reproducible, and automated pipeline to collect raw data files (textual, visual, and acoustic), extract multimodal features, and release them. % embeddings.
% In addition, we release the mappings with URLs to the raw data files, as well as the multimodal features we extracted using several state-of-the-art encoders. 
Our qualitative and quantitative analyses confirm that all the modalities are worth considering and provide complementary information.
We plan to maintain our resource by regularly checking whether any videos or posters are added or removed and updating the mappings accordingly. In addition, we plan to include additional encoders for each modality and are open to collaborating with the community to increase the availability of our resource.

\begin{acks}
{This paper has been accepted at RecSys 26 and is published in \emph{Proceedings of the 20th ACM Conference on Recommender Systems (RecSys '26)}, https://doi.org/10.1145/3773078.3831852}
\end{acks}

\bibliographystyle{ACM-Reference-Format}
\bibliography{paper_biblio}

\end{document}